\documentclass[aps,prl,showpacs,%twocolumn,
amssymb]{revtex4}
\usepackage{CJK}

\usepackage[usenames,dvipsnames]{color}

\usepackage{ulem}
\usepackage[T1]{fontenc}
\usepackage{calligra}

\usepackage{graphicx}% Include figure files
\usepackage{dcolumn}% Align table columns on decimal point
\usepackage{bm}% bold math
\usepackage{times}
\usepackage{verbatim}
\usepackage{graphicx}
\usepackage{graphics,epsfig}

\usepackage{theorem}
\usepackage{makeidx}
\usepackage{amsmath}
\usepackage{epic}
\usepackage{amscd}

\usepackage{bbm}
\usepackage{xy}
\usepackage{amssymb}
\usepackage{epsfig}

\usepackage[caption=false]{subfig}

\def \phi {\mbox{$\varphi$}}

\begin{document}

%\begin{CJK*}{GB}{}
%\begin{CJK*}{KS}{}

\title{Disorder and Power-law Tails of DNA Sequence Self-Alignment Concentrations in Molecular Evolution}
\author{Jian-Zhou Zhu{$^{1}$\footnote{zhujianzhougml@gmail.com}
, Kun Gao{$^{2}$}\footnote{maplebridger@gmail.com}
, HongGuang Sun{$^{1,3}$}, }%\footnote{Correspondence Email: jz@sccfis.org}
}
\affiliation{{$^1$}Su-Cheng Centre for Fundamental and Interdisciplinary Sciences, Gaochun, Nanjing 211316 China
and Li Xue Center, Gui-Lin Tang Lab., 47 Bayi Cun, Yong'an, Fujian 366025 China\\
{$^2$} Physics and Biology Unit, Okinawa Institute of Science and Technology (Graduate University), Okinawa\\
$^3$College of Mechanics and Materials, Hohai University, Nanjing 210098, China
}

\begin{abstract}
The self-alignment concentrations, $c(x)$, as functions of the length, $x$, of the identically matching maximal segments in the genomes of a variety of species, typically present power-law tails extending to the largest scales, i.e., $c(x) \propto x^{\alpha}$, with similar or apparently different negative $\alpha$s. Recently, the stick breaking phenomenology for the mutation effect on the duplicated segment has been proposed to address such tails. We recognize that randomness is intrinsic to the molecular evolution system at different levels. By introducing frozen randomness in the setup (the mutation rate $\mu$, the initial condition and/or the input) of a fragmentation model for the dynamics of the concentration, we obtain solutions $\propto x^{\alpha}$ for $x\to \infty$, time-dependent or not, which is in contrast to the only steady power-law solution $\propto x^{-3}$, for $x\to 0$, of the pure model (without disorder).
We also present self-alignment results showing more than one scaling regimes, consistent with the theoretical prediction from the existence of more than one algebraic terms which dominate at different regimes.
\end{abstract}


\maketitle

\begin{figure}[h!]
\includegraphics[width=0.45\textwidth]{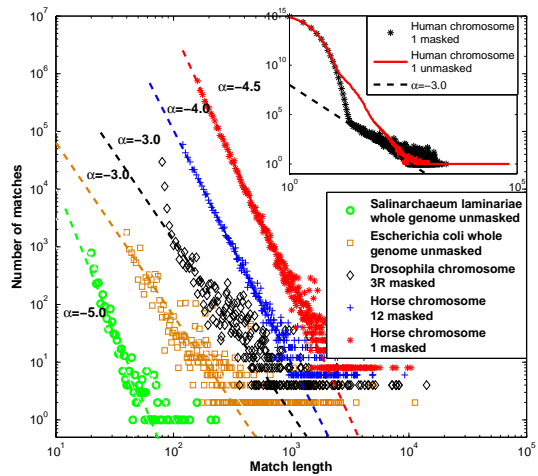}
\caption{Concentrations, computed by MUMmer \cite{MUMmer} and plotted in log-log scales, of maximal matching segments in the self-alignments of the genome sequences of various species, shifted apart for better visualization, showing similar or apparently distinct scalings extending to the largest scales: All data, except those in the inset whose plot of human genome reproduces with Chromosome 1 the results of Massip and Arndt \cite{MassipArndt13}, are shown only for scales above 20bp below which there is no power law and where the computation is also very expensive.
%The scaling ranges for procaryotic species usually lie inbetween a couple of tens and several hundred bp, while those of eucaryotic ones are in between several tens to a couple of kbp.
Eucaryotic species, unlike the procaryotic ones, are in general seriously affected by simple repeats as shown by the data of homo sapiens (human) where the very small scale range is also presented to show that the data is basically exponential, $\propto (1-p)^2p^{x-2}$ \cite{MassipArndt13}, due to random matching from the finite number $1/p$, now 4 for ``A'' ``T'' ``C'' ``G'', of the alphabets and coagulation effects (also exponential distribution with the same equal probability - $p$ - assumption). Dashed lines are exact scaling laws for reference.
%Except for the inset, all other alignment concentrations are only plotted for scales larger than 20bp below which are basically the exponentials.
}\label{fig:powerlaw}
\end{figure}

%\section{Introduction}
%\textit{Introduction.}---
The effects of duplication and mutation responsible for the `complexity in genomes' (\cite{Cgenomes} and references therein) are crucial for evolution, such as the generation of biodiversity (e.g., Ref. \cite{EvoAftDup} for an overview of theory and mathematical models along with practical examples). Compared to other processes such as recombination, the dynamics of duplication is different, thus it is helpful to isolate its fingerprints in the genome sequences, say, by masking simple repeats \cite{repeatmasker}, from the data for separate studies such as the neutral evolution dynamics, among the various debatable considerations (cf. Ref. \cite{NatureDebate} for a recent dialog.) One can also try to obtain information about life concerning disease susceptibility and paralog v.s. ortholog issues etc., from studying the duplication and mutation (see, e.g., \cite{BaileyEichler06,Gao14}).
Fig. \ref{fig:powerlaw} presents some examples of the concentrations (histograms) $c(x)$s, of the maximal exactly matching nucleotide(-pair) segments of deoxyribonucleic acid (DNA) double helixes, as functions of the match length $x$, from both eucaryotic and procaryotic species. The plots in log-log scales suggest power-law tails $c(x)\propto x^{\alpha}$ extending to the largest lengths/scales. We see that the scaling exponents can be similar or apparently different for a variety of species; the power laws may even not be the same for two distinct chromosomes of the same species (c.f. more details in the caption.)
Such a phenomenon raises questions such as whether there are some universal mechanisms behind them? And what they mean for genomic physics? Massip and Arndt \cite{MassipArndt13} took the exponent of the repeat-masked whole human genome sequence to be exactly $-3$, a typical value for some specific chromosomes of various eukaryotic species \cite{gao_miller,taillefer_miller}. They also showed that the repetitive elements \cite{repeatmasker} greatly deteriorate the scaling law. Li et al. \cite{LiETC14,Li14} recently also discovered other relevant forms of power-law distributions. It thus appears to us that a reasonable model, especially that from the null hypothesis of neutral molecular evolution, should present the power-law tails at large scales as well as the different possible scaling exponents with a common or somewhat ``universal'' mechanism among them.

%To address the case of $\alpha=-3$,
Ref. \cite{MassipArndt13} further introduced an analytically trackable pure fragmentation model with the stick breaking phenomenology for the point mutation effect on a duplicated segment (Koroteev and Miller \cite{koroteevmiller2011} had done simulations with descriptive procedures containing some of the essential features, and they later presented similar studies \cite{KoroteevMiller13}.)
The scenario was the following: The matching segments, defined by copies of nucleotide(-pair) sequence that are the same but are different when extended beyond either end, come from segmental duplications subject to point mutations; mutations ``break'' the matching segments into pieces of shorter matching segments% (Step 2 in Fig. \ref{fig:cf})
. By suitably assigning the mutation rate per site, $\mu$, and (linearly) balancing the gain and loss at each scale, they obtained the fragmentation model for the evolution of the concentration whose solutions happen to be known \cite{zm2,ben-naim}. For example, the steady-state solution to the model with input scales as $x^{-3}$  for the ``head'' with $x\to 0$ \cite{ben-naim} (compared to the system size, say), instead of the ``tail'' ($x\to \infty$ compared to the number of alphabets, say.) The data may indicate a power-law `tail' rather than `head' (we suggest such terminologies for discrimination). Although the notions of being `large' compared to some small scale and being `small' compared some large one do not directly conflict, we will show that physics of the scaling laws are different.
%\begin{figure}[h!]
%\begin{center}
%  % Requires \usepackage{graphicx}
%  \includegraphics[width=0.45\textwidth]{cf1.eps}
%  \caption{The example of duplication and mutation, and, the fragmentation and coagulation effects.}\label{fig:cf}
%\end{center}
%\end{figure}

%\subsection{Fragmentation and coagulation effects from duplication and mutation}
Let us %go with a more technical introductory discussion and
call an $x$-length segment $x$-matching or $x$-unmatching, depending on whether it belongs to the set of matching segments of length $x$ %, with concentration/distribution $c(x)$,
or not. %First of all, we note that a
Beside fragmentation, a mutation may have %two-fold effects on matching segments:
%\begin{itemize}
%  \item
%One obvious effect is fragmentation, i.e., an $x$-matching segment is broken into shorter matching segments;
%  \item
%the opposite effect, not discussed previously, is
coagulation effect, i.e., a point mutation may unite its two sides to become a longer match.
Yet another ``null'' effect is that the mutation turns an $x$-matching segment into another $x$-matching one. %, without changing the number/concentration of the $x$-matching segments.
The concentration (Fig. \ref{fig:powerlaw}) results from such effects.
It is also easy to recognize that in the detail molecular dynamics, the duplication appear stochastic in position and time; and, similarly does mutation: Such is called a `low-level' or `fundamental' randomness, which of course can be further detailed and divided into even lower level or more fundamental randomness. % (concerning geometry and content etc.)
At a `high' or `phenomenal' level, the input/duplication rate at scale $x$ of the concentration dynamics may randomly depend on time or even nonlocally in the sense of scale on the concentration $c(y)$; and, again, similarly for mutation. So, in general the fragmentation-coagulation effects for $c(x,t)$ evolution can be described by a stochastic nonlinear integro-differential equation (for coagulation-fragmentation models, see, e.g., \cite{Broizat10} for kinetic descriptions analyzed rigorously already, and references therein.) %Though already very intuitive, i
It may be useful to be a bit more definite on a possible origin of the disorder at such a `high' level: %we will be discussing
%which is that t
There are many different segments of the same matching length; or, in other words, an $x$-matching set contain segments of different local (arrangement of) nucleotide base pairs and/or different ``ribbon'' writhes, torsions and twists of the double helixes (the structures of DNA). %: To get the general idea, see, {\it e.g.}, Chap. 4 of \cite{WatsonETCbook} for both structure and topology.
%even with the same sequence and length (may be called \textit{topoisomers} segments) constraints.
These $x$-matching but different segments characterize mutations and/or duplications differently. The (random) environments also add to the disorder in the duplications and mutations.
Since we can not or need not know exactly all the details, specific statistical distribution is applied.
%For this, as an idealization, one can think of a big ensemble containing a distribution of sub-ensembles, or, in other words, all the $x$-matching segments do not correspond to independent-identical-distribution (i.i.d.) variables but contain (infinitely) many subsets of elements corresponding to i.i.d. random variables.
A single realization or simple mean-field treatment of the $c(x,t)$ dynamics is not sufficient. A systematic derivation of the random forms of mutation, duplication and coagulation (in the reaction rates \cite{Broizat10}) etc. however has not been available. The molecular biology considerations and the accumulated vast amount wisdom about coagulation-fragmentation processes provide useful clues for us to proceed tentatively. For example, given the mutation rate, the fragmentation effect may well be modeled by the conventional fragmentation model; as for the coagulation, due to the fact that the number of alphabet, 4, in genome sequences is very small compared to the total length, its effect should mostly concentrate at small scales (actually the coagulation effect and the finite-alphabet effect are not completely separated). Thus, as an application and development, it is reasonable to follow Ref. \cite{MassipArndt13} to start with the simple one \cite{zm2}:
\begin{equation}
\label{rate}
{\partial c(x,t)\over \partial t}=-\mu xc(x,t)+2\int_x^\infty \mu c(y,t) dy + f(x),
\end{equation}
where, compared to previous studies, the new element in the model lies in the stochasticity of the initial condition $c(x,0)$, of $\mu$ and of the input $f$. [For convenience, we are different to Ref. \cite{MassipArndt13} with a factor of $2$ (absorbed into the mutation rate $\mu$).] And, the input $f$ is now used to model the gross contributions from the duplications and the (nonlinear) coagulation effects of mutations. $\mu$ is assumed to be independent of $x$.
%The solutions to the deterministic Eq. (\ref{rate}) has been known \cite{zm2,ben-naim}.
Continuously forced, the final steady solution [Eq. (\ref{cinfty}) below], without time dependence requires mono-scale input to have $c(x)\propto x^{-3}$ extend to largest scales. Such `monodispersion' \cite{ben-naim} however results in a pulse and truncation at the input scale $x=K$; thus, to our point of view, a dilemma calling for alternative treatments and interpretations \footnote{Time integration result of the ``decaying'' `monodispersion' [c.f., Eq. (\ref{monodispersion}) below] from $t=0$ to $\infty$, thus killing the time dependency, has been further elaborated by F. Massip, M. Sheinman, S. Schbath and P. F. Arndt (``How Evolution of Genomes is Reflected in Exact DNA
Sequence Match Statistics'', MBE electronic publication ahead of print), who argued yet another $x^{-4}$ result for the retroduplication model, but for $x\ll K$.}.
%Actually time accumulations of ``history'' ending at infinite time $+\infty$ and reaching the current moment $t$ are given precisely by the steady-state and general time-dependent full solution, respectively, to be presented below.
\textit{To avoid dealing directly with the stochastic integro-differential equation, we further apply frozen randomness. % in the mutation rates and inputs.
Our disorder realizations are specified by the distributions of the parameters in the ansatzes of the initial conditions and of the inputs, and by the distribution of $\mu$, which turns out to not only resolve the `dilemma' but also offer other results such as a spectrum of exponents as the data suggest, time dependency and extra power-law regime(s).}

%\section{Solutions}
\textit{Solutions.}---
%It is not completely clear whether we should best treat our present genome data as the ``decaying'' state, given the duplications happened a long time ago and the on-going mutations, or as the statistical steady state, balanced among the effects of duplications and mutations, or most generally a state corresponding to the time-dependent solution with initial data and continuous input. The answer probably depends on the time scale we want to put our current observation in.
The general time-dependent full solution of course is the most meaningful, but we will examine also the ``decaying'' and ``steady-state'' solutions which actually underline the basic features of the general full solution.
%

%\subsection{``Decaying'' solution}
\textit{``Decaying'' solution.}---
Let's start with the \textit{``decaying'' case} with $f(x)=0$ which may correspond to the case dominated by the fragmentation process.
For $c(x,0)=\delta(x-K)$, the `monodispersion' solution is \cite{zm2,MassipArndt13}
\begin{equation}\label{monodispersion}
c(x,t)=e^{-K \mu t}\delta(x-K)+[2\mu t+(\mu t)^2(K-x)]e^{-x\mu t}
\end{equation}
for $0 < x\le K$, otherwise null; and, in general
\begin{equation}\label{decaying}
    c(x,t)=e^{-x\mu t}\big{\{} c(x,0)+\int_x^{\infty} c(y,0)[2\mu t+ (\mu t)^2(y-x)]dy\big{\}}.
\end{equation}
Note that the word ``decaying'' does not indicate any dissipation of the mass, $M=\int_0^{\infty}xc(x,t)dx$, which is conserved \cite{zm2}.

\begin{figure}[h!]
  \includegraphics[width=0.45\textwidth,]{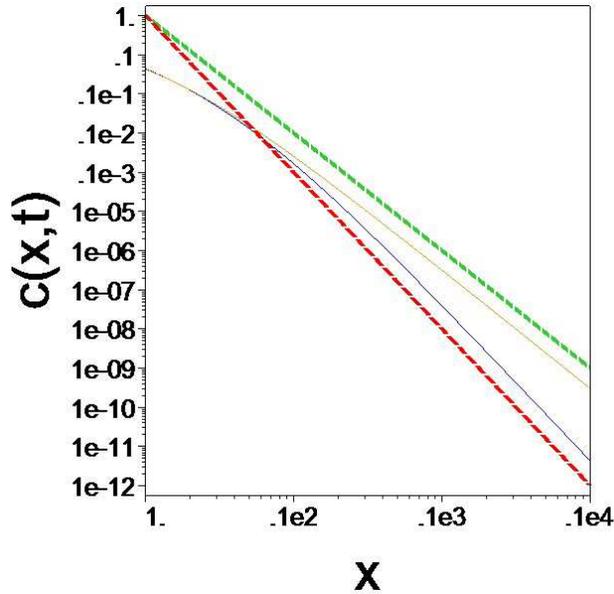}\\
  \caption{The curved lines are the solutions with asymptotic tails $x^{-3.0}$ and $x^{-4.0}$ as are the straight reference lines.}\label{fig:decaying}
\end{figure}

We first check an example with a realization with $\tilde{c}(x,0)=e^{-\tilde{s}x}$, as also in Ziff and McGrady \cite{zm2}, but $\tilde{s}$ has quenched disorder: We always use a tilde to denote a realization of the disorder.
Our solution can be obtained in two steps. First, fix $\mu=\tilde{\mu}$ %and use the replacement of $t \to \tilde{\mu} t$ to Ziff and McGrady's \cite{zm2} Eq. (11)
to get
\begin{equation}
\label{csnapshot2}
\tilde{c}(x,t)= \frac{(\tilde{\mu} t+\tilde{s})^2}{\tilde{s}^2} \exp\{- (\tilde{\mu} t+\tilde{s}) x \}.
\end{equation}
Then we integrate over the distributions $P_{\tilde{\mu}}$ of $\tilde{\mu}$ and $P_{\tilde{s}}$ of $\tilde{s}$ to get the final averaged solution. The more general result in some appropriate conditions is possible to be evaluated with Laplace's method \cite{BenderOrszag}. We illustrate them with definite examples as follows. 
For instance, assuming $$P_{\tilde{\mu}}(\mu)\propto \mu^n e^{-\lambda \mu} \ \text{and} \ P_{\tilde{s}}(s) \propto s^m e^{-\Lambda s},$$ with $n\ge 0$ and $m>1$ (for convergence of the continuous integrations), gives for $x\to\infty$ and $t\to\infty$ asymptotically
\begin{equation}\label{eq:decaying}
   c(x,t) \propto x^{\alpha_d}t^{\beta}: \ \alpha_d=-(n+m+2) \ \text{and} \ \beta=-(n+1).
\end{equation}
The restriction $m>1$ here is due to the requirement of the convergence of the continuous integral, and, in practice, with discrete and/or finite scales, it may be relaxed by easily tuning the ansatz at small argument or controlling the integration range. 
Fig. \ref{fig:decaying} shows, with $\lambda=2.0$, $\Lambda=5.0$ and $t=2.0$, a typical plot of solutions for $n=0$, $m=1.0$ and $2.0$, compared to lines with exact slopes $-3.0$ and $-4.0$: As said, for $m=1.0$ the integration diverge at $s=0$, so we obtain the result approaching $x^{-3.0}$ in the figure by setting $P_{\tilde{s}}(s)=0$ at very small $s$.
We just remark that the above result turns out to be quite robust for a large class of reasonable ansatzes: %; for instance, using $P_{\tilde{\mu}}(\mu)\propto \mu^n e^{-\lambda \mu^2}$ still produces the above algebraic tails.
One may still get algebraic tails with other ansatzes, such as
$$c(x,0)=x^p e^{-\tilde{s}x^2}, \ P_{\tilde{\mu}}(\mu)\propto \mu^n e^{-\lambda \mu^2}, \  P_{\tilde{s}}(s)\propto s^m e^{-\Lambda s^2};$$
$$c(x,0)=\frac{x^p}{(x+\tilde{s})^{q}}, \ P_{\tilde{\mu}}(\mu)\propto \frac{\mu^n}{ (1+\mu)^{r} }, \ P_{\tilde{s}}(s)\propto \frac{s^m}{ (1+s)^{z} }.$$
%and so on and so forth.
%The asymptotic time behavior is also algebraic, but the value of $\beta$ also depends on the choice of the ansatzes.
So, the algebraic tails appear to be the generic output of the combination of such distributions.
%Mathematically these results remind us but are different to the ``weighted mixtures'' of Willinger et al. \cite{Willinger04} who requires the distribution (our concentration after normalization) be a scaling function. But, indeed now the power-law tail is also what they stated to be ``more normal than normal [Gaussian]''.

%After discovering the generic mechanism, we need to identify the details in our specific problem.
In genome sequence, the question is then what exactly are the initial distribution, the disorder in it and in the duplication and mutation rates? The state-of-the-art of biophysics (theory and documented data) can not give satisfying answers, but we tend to believe that the above ansatzes, with possible quantitative modifications, used for the explicit calculations should be qualitatively `reasonable' in describing what has been happening in nature, since they just simply represent the obvious facts of peaks at some (moderately) small values and the convergence properties. But all these follow the assumption of quenched disorder. Quenched disorder should be considered to be a working hypothesis or an effective modeling strategy. %How reasonable or close it is to the nature needs to be checked against the data.
Intuitively, being frozen in time of the disorder in the initial concentrations sounds natural, but that of the mutation rates is just a working simplification.

%\subsection{Steady state solution}
\textit{Steady state solution.}---
The final steady-state solution with a realization of input reads \cite{ben-naim}
\begin{equation}
\label{cinfty}
\mu \tilde{c}_\infty(x)=x^{-1}\tilde{f}(x)+2\,x^{-3}\int_x^\infty dy\,y\tilde{f}(y).
\end{equation}
So, whatever the steady input is, a tail of $-3$ for $x\to \infty$ is not consistent: One can just check situations with $\tilde{f}(x)$ decaying as, faster than or slower than $x^{-2}$. Such a model can produce genuine slope steeper than  $-3$ only with an input of power law steeper than $-2$: From Eq. (\ref{cinfty}),
if and only if $\epsilon >0$,
the input of slop $-2-\epsilon$ produces a distribution of slope  $-3-\epsilon$. For instance, an exponential input gives an exponential tail with algebraic prefactor. Note in particular that $\epsilon$ cannot be 0. Such observations were already partly made earlier semi-empirically \cite{koroteevmiller2011}.

The restriction of power-law input for a genuine power-law tail is removed by disorder: For example, an input $\tilde{f}(x)=\tilde{\lambda} e^{-\tilde{\lambda} x}$ gives
$$\mu \tilde{c}_{\infty}(x)={\frac {2\,\tilde{\lambda}\,x+{\tilde{\lambda}}^{2}{x}^{2}+2}{{\tilde{\lambda}}{{\rm e}
^{\tilde{\lambda}\,x}}{x}^{3}}}$$
which, when $\tilde{\lambda}$ has disorder (quenched) of a distribution ansatz
$$P_{\tilde{\lambda}}(\lambda)\propto \lambda^n e^{-\Lambda \lambda} \ \text{with} \ \Lambda>0 \ \text{and} \ n>0,$$
produces an asymptotic tail ($x\to\infty$)
\begin{equation}\label{eq:steady}
c_{\infty}(x) \propto x^{\alpha_s}: \ \alpha_s=-(n+3).
\end{equation}
The exponent $\alpha_s$ is independent of $\mu$ disorder.
%, which, as we will see later, however is important for the algebraic tail in the full time-dependent solution, just as in the decaying solution.}
Just as the ``decaying'' case, the condition $n>0$ is for convergence of the continuous integration, and in practice with discrete and/or finite scales, this condition may be relaxed by tuning the ansatz at small arguments; and, also, some other `reasonable' (again, in the sense of consistent with our understanding of the duplication and coagulation effects from mutations) ansatzes for the input also produce the power-law tails.

\begin{figure}[h!]
\centering
  \includegraphics[width=0.45\textwidth]{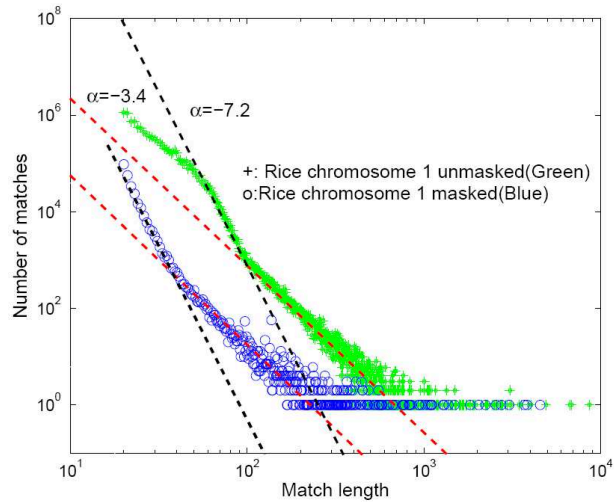}\\
  \caption{The concentrations of rice chromosome $1$ appear to have two power-law regimes. Dashed lines with exact power laws are given for reference.}\label{fig:rice}
\end{figure}

%\subsection{Full time-dependent general solution}
\textit{Full time-dependent general solution.}---
It seems most natural to consider the general time-dependent solution for interpretation and prediction.
As already given with the Mellin transform and Charlesby method by Ben-Naim and Krapivsky \cite{ben-naim} whose details we resist to reproduce here, the solution is simply the linear superposition of the previous decaying solution and the time-dependent solution with input but without memory of the initial condition, with a similar structure of the combination of the ``decaying'' and steady-state solutions; thus from Eqs. (\ref{eq:decaying}) and (\ref{eq:steady}) we immediately conclude, with two dominant representative algebraic terms in $t$ and $x$, for large $x$ (and $t$ in the first term): %, which ends exhausting all three possibilities in a satisfying way.
\begin{equation}\label{eq:full}
    c(x,t)\propto C_1 t^{\beta} x^{\alpha_d} + C_2 x^{\alpha_s}.% + C_3 t^{-\beta_3}x^{-\alpha_3}.
\end{equation}
We remark that at different scales separated far apart, the two components may dominate respectively. Actually we have already neglected subdominant algebraic terms for $x\to\infty$ and/or $t\to\infty$ while giving the final decaying or steady-state solution, but the subdominant algebraic term(s) could dominate at some intermediate regime(s), showing different power law(s): Whether or not this is happening in the data depends on the coefficients, $C_i$ whose determination is a combination of the pure fundamental fragmentation(-coagulation) dynamics \cite{MassipArndt13} and nature's setup of the disorders. For example, the alignment concentrations of rice, as given by Fig. \ref{fig:rice}, appear to support two power-law regimes, especially for the unmasked data. Some other data also present such a similar feature, though do not always have as sharp results \cite{Shallower}. On the other hand, as a `negative' effect, such mixed algebraic components may also add to the ambiguity in detecting the power law when they don't separate well or the data don't have enough range of scale to separate them \cite{gao_miller}.

%\section{Discussion}
\textit{Discussion.}---
%We have shown that disorder in the fragmentation(-coagulation) dynamics resolves the dilemma concerning `monodispersion' in getting the power-law tails reaching the largest scales from a single realization or simple mean-field treatment. It may also be more appropriate to regard the current concentrations as still evolving, i.e., depending on time. Such disorder enables the null hypothesis to be open to various add-ons not precisely known.
Various ansatzes of disorder have been found to produce power-law tails in quite a generical way, which on the one hand demonstrates some universality in nature under this theoretical framework, while on the other hand they have interesting differences in biophysics.
The results points to the directions of measuring the disorders in the data, which requires tremendous experiments/computations of genomes to quantify the various ingredients such as the distributions of the duplication, the mutation rates, the coagulation effects, which will help to narrow down our somewhat general results to even more specific biophysics and will help preparing for the full stochastic dynamical modeling.%\textbf{: We can write one realization solution of Eq. (\ref{rate}) with the input being time dependent, denoting the ``decaying'' solution [Eq. (\ref{monodispersion})] to the mono-scale (at $x_0$) initial (at time $t_0$) condition problem as $C(x,t;x_0,t_0)$, in the form $\tilde{c}(x,t)=\int C(x,t;y,t_0)\tilde{c}(y,t_0)dy + \int_{t_o}^t \int C(x,t;y,s)\tilde{f}(y,s)dyds$, and $c(x,t)$ would be the ensemble average of $\tilde{c}$ over the initial condition and the input; if the input is not statistically steady, the solution would never reach a steady state. Specification of such a solution requires further documentation of data}.

It is also interesting to identify what is specific (especially concerning disorder) to the masked repeats and the possible relevance with selection with regard to our Fig. \ref{fig:rice}: To our best knowledge, so far there has not been well established selective mechanisms that can be claimed to be in favor of or against a power-law concentration. And, there are other models of molecular evolution emphasizing on different aspects of observations (cf., Refs. \cite{LiKaneko,MesserArndt,Stanley,Grosberg,Cgenomes} for a partial list), and it is hoped that some of the considerations here may also be applicable to strengthen some of them.

%\section{Acknowledgement}
This work benefited from the helpful discussions with J. Miller's group, and we are indebted to W. Li and P. F. Arndt for kindly explaining their fresh publications.

%\end{CJK*}
%\end{CJK*}

\end{document}